\documentstyle[12pt,aasms4]{article}
\def\gtsim{\lower.5ex\hbox{$\; \buildrel > \over \sim \;$}}
\def\ltsim{\lower.5ex\hbox{$\; \buildrel < \over \sim \;$}}

\slugcomment{}

\lefthead{Mushotzky \& Scharf} \righthead{The Luminosity-Temperature
Relation at z=0.4 for Clusters of galaxies}

\begin{document}

\title{The Luminosity-Temperature Relation at z=0.4\\
    for Clusters of galaxies}

\author{R.F. Mushotzky and C.A. Scharf\altaffilmark{1}}
\affil{Laboratory for High Energy Astrophysics, Code 662, NASA/Goddard
Space Flight Center, Greenbelt, MD 20771, USA}

\altaffiltext{1}{also, Dept. of Astronomy, University of Maryland, College
Park, MD 20742-2421, USA}

\begin{abstract} We have obtained the first large sample of accurate
temperatures for clusters at $z>0.14$ from ASCA. We compare the luminosity
temperature (L-T) distribution for these clusters with the low redshift
sample of David et al (1993) and find that there is no evidence for
evolution. We also find that the intrinsic variance in this relation is
roughly constant with redshift. Additionally, there is no detectable
change in the relationship of optical velocity dispersion to X-ray
temperature with redshift.  Most cosmological simulations driven primarily
by gravity predict substantial changes in the L-T relation due to the
recent rapid growth of clusters. Our results are consistent either with
models in which the cluster core entropy is dominated by pre-heating, or
with low $\Omega$ models in which cluster structure does not evolve
strongly with time. The intrinsic variance in the L-T relation at a fixed
redshift can be due a variety of possibilites e.g. a change in the
baryonic fraction from cluster to cluster, variation in the fraction of
the total energy in the system arising from shocks or supernova heating or
variations in the emission measure distributions in multiphase gas. 

\end{abstract}

\keywords{galaxies:clusters:general - X-rays:galaxies -
cosmology:observations }

\section{Introduction} 

Clusters of galaxies are the largest relaxed systems in the universe and
as such provide a strong test of theories of the origin of and evolution
of structure. A critical measurement (\cite{fre90}) is the relationship of
the mass of a system to its temperature. This can be best probed by
comparing the evolution of the X-ray luminosity (which is related to the
baryonic mass) to the x-ray temperature, which is roughly proportional to
$M^{2/3}$. Calculations of the evolution of these quantities (c.f.,
~\cite{kan94}, \cite{kai91}) indicate that, in a closed universe, where
gravity dominates the evolution of the gas and the dark matter, the
relationship between luminosity and temperature should evolve strongly at
all redshifts, with significant changes occurring at $z<0.5$. However the
nature and amplitude of the evolution is dependent on the details of the
cosmological model (\cite{bry94}). 

For example, evolution is less in the CDM+ $\Lambda$=1 model than in a
pure CDM universe and is also less in a low density universe, or one in
which the entropy of the gas is relatively large at high redshift
(\cite{kai91}, \cite{bow97}, \cite{nav95}). This evolution is not
necessarily monotonic. Cen \& Ostriker (1994) show that in a
CDM+$\Lambda$=1 model there is an almost constant mean cluster temperature
from $z\sim 0.3$ to the present while in a CDM model there is a sharp
decrease in the temperature from $z=0.0$ to $z=0.3$. 

At low redshift the L-T relation is well measured (\cite{mus84},
\cite{edg91}, \cite{dav93}). While the correlation is good, there is a
fair degree of scatter (\cite{fab94}), with cooling flow clusters being
rather more luminous for a given temperature. The theoretical prediction
for the relationship between luminosity and temperature in a CDM universe
dominated by gravity (\cite{fre90}, \cite{kan94}, \cite{evr90}) is
somewhat flatter than observed. There are various possible explanations
for this such as systematic variations in the baryonic fraction with
cluster mass (\cite{dav93}) or pre-heating (\cite{evr90}). 

At low redshifts the relationship between the cluster velocity dispersion
($\sigma_V$)and the gas temperature (\cite{bir95}, \cite{bac94},
\cite{gir96}) is very tight, consistent with both tracing the same
potential. This strong agreement is not generally predicted to continue at
higher redshift. 

ASCA data allow, for the first time, a measurement of the L-T and the
$\sigma_V$-T relationships at $z\gtsim 0.1$ from a large sample of
clusters. Thereby placing strong constraints on theories of cluster
evolution.  In this paper we use a sample of 38 clusters of galaxies from
$z\sim 0.14-0.55$ obtained from the ASCA archives. This sample, while not
homogeneously chosen, is large enough to constrain this relationship at
$\left<z\right>\sim 0.3.$

The only previous attempt to measure the L-T relationship at $z\gtsim 0.1$
(\cite{hen94}) was consistent with no evolution, but with rather large
errors.  However, this study relied on summing a large number of low
signal to noise data sets from the flux limited Einstein medium survey and
was therefore fundamentally different in character. The ASCA data have
well determined temperatures for each cluster but are drawn from a
heterogeneous sample. 

\section{Observations}

We have extracted the spectra from 38 clusters of galaxies at $z>0.14$
from the ASCA archive (Table 1). The sample simply consists of all
clusters available before Nov. 1996 for which the data are publicly
available, the temperature uncertainty is $ \delta T/T\ltsim0.3$, and
there is no strong substructure visible in the ASCA image.  In our
analysis we allowed the cluster abundance and galactic absorption to be
free parameters.  For clusters at $z\sim 0.3$ this effectively requires
cluster flux $\times$ $\sim 4\times10^{-8}$ ergs cm$^{-2}$. 

Examination of the data (Figure 1) shows a relatively strong selection
bias against clusters of $L_{bol}< 10^{45}$erg s$^{-1}$ at $ z \sim 0.3$.
This comes from the desire of many of the proposers to obtain higher
quality spectra, requiring bright sources and/or longer exposures than
normal with ASCA, combined with the relatively small number of known
clusters in this luminosity/redshift range. 

At $z \sim 0.3$ clusters of $L_{bol}< 10^{45}$ are below the threshold of
the ROSAT all sky survey and mainly come from the EMSS data base
(\cite{gio90}) and serendipitous ROSAT sources. At flux levels $\gtsim
10^{-12}$erg s$^{-1}$cm$^{-2}$ the ASCA data analysis is relatively
straightforward (see~\cite{mus97} for a discussion of the higher signal to
noise objects in this sample). We have used the latest calibrations
including the gain change in the SIS detectors. 

Quoted errors are 90\% confidence for one parameter ($\chi^2$+2.71).  We
use $q_{0}=0$ and $H_{0}=50$ km s$^{-1}$ Mpc$^{-1}$ in this paper. Our
derived temperatures agree very well with those obtained by~\cite{tsu96},
for common objects and the previously published results of~\cite{sch97},
\cite{all96b}, \cite{bau94} and~\cite{don96}. 
 
We have integrated the spectra over 3-6 arcmin in radius, depending on
cluster redshift, in an attempt to get the average spectrum in a fashion
similar to those obtained by non-imaging proportional counters for lower
redshift objects. Thus our values are directly comparable to those of the
non-imaging experiments. The effects of cooling flows (\cite{all96b}),
mergers (\cite{hen96}) or non-isothermal profiles (\cite{mar96a}) can
change the mean temperature by up to $\pm$20\%. For example, in A1835
which has a very large cooling flow, the mean kT=8.15 keV (Table 1
and~\cite{all96a}) changes to 9.5 keV with the inclusion of a cooling flow
in the spectral fit. The uncertainties in the temperature for these high
$z$ systems are similar to or less than the errors obtained from EXOSAT
(\cite{edg91}), HEAO-1 (\cite{mus84}) or the Einstein MPC (\cite{dav93})
proportional counters, and thus form a good comparison sample. We have
added a few low $z$ clusters to the David et al (1993) compilation (A3158
and A3581) and obtained accurate ASCA temperatures from our own analysis
and the literature for those low $z$ clusters with large temperature
uncertainties in the David et al (1993) compilation (A399, A401, A3112 and
A3391) to obtain an essentially flux limited low $z$ sample resulting in a
total sample of 102 clusters from $z\sim 0.01$ to $z\sim 0.55$. 

\section{Analysis}

Comparison of this sample ($z>0.14$) with the enhanced David et al (1993)
sample of low redshift clusters ($z<0.1$) shows no evidence for a change
in the L-T relationship (Figure 1) over the whole luminosity range covered
($\log_{10}L_{bol}>44.3$). However the ASCA data show a strong bias at the
low luminosity end of the distribution due to the absence of objects in
the database. We believe this selection effect accounts for the apparent
flattening of the L-T relationship for the high $z$ clusters at low
luminosities (Figure 1). Thus we restrict our analysis to $45.2\leq
\log_{10}L_{bol} \leq 45.7$ for which there are no strong selection
effects and for which there is maximum overlap of the low and high $z$
data sets and a reasonable set of objects (39).  Applying the two
dimensional Kolmogorov-Smirnov test we find that at $45.2\leq
\log_{10}L_{bol} \leq 45.7$ the low and high $z$ data show no evidence of
being significantly different (probability of being different $<60$\% for
all high/low redshift divisions). The inclusion of the lowest luminosity
points, combined with a smaller data set led Tsuru et al (1994) to infer a
change in the L-T relation which is not seen in the present data. 

The high $z$, large accretion rate, cooling flow clusters in Figure 1
(e.g.  RXJ1347, A1835, EMSS1455 occupy a ridge line to the right in the
L-T plot. This property was noted in the low $z$ sample by Fabian et al
(1994). 

We estimate the allowed change in either the normalization or slope of
this relation by examining the variation in the mean temperature in
redshift shells.  In Figure 2 the mean cluster temperatures determined by
maximum likelihood are plotted as a function of redshift bin.  Much of the
plotted allowed range in the mean temperature is not due to uncertainties
in the data but to real width in the L-T relation, as seen in the low $z$
sample (Figure 1 and~\cite{fab94}). Solid vertical error bars represent
the 90\% confidence limits on $\left<T\right>$, dotted error bars indicate
the intrinsic dispersion of the $L_{bol}$-$T$ distribution estimated by
likelihood analysis. Using $q_{0}=0.5$ gives a small but significant
change in the normalization, such that higher redshift clusters are less
luminous for a given temperature.

The limits on temperature evolution are estimated from Figure 2 as:
$\Delta \log_{10}\left< T \right>\simeq 0.04$ for $\Delta
\log_{10}(1+z)\simeq 0.15$ at fixed luminosity. 

We also compare the relationship of velocity dispersion to temperature and
luminosity, using the large sample of Fadda et al (1996) for the low $z$
sample and the Carlberg et al (1996) and Fabbricant et al (1991) data for
the high $z$ sample. We find that the high $z$ clusters in the temperature
range 4-9 keV show virtually identical $\sigma_V$-T and $\sigma_V$-L
relationships to those of the low $z$ sample (Figure 3). AC118 and A1689
are clearly discrepant, indicating non-virial velocity dispersion.

\section{Discussion}

The lack of evolution in the L-T relation at $z<0.5$ combined with the
recent upper limits on the evolution of the luminosity function over the
same redshift range places strong constraints on all models of cluster
evolution (\cite{kai91}). As pointed out in detail by Navarro, Frenk and
White (1995) a model in which the intial specific entropy is large results
in little, if any, evolution in the L-T plane and better matches the
overall slope of the observed L-T relation. Metzler and Evrard (1994)
describe a model in which entropy due to the creation of the metals is
explicitly included, increasing the overall entropy of a kT$\sim$ 6 keV
cluster by a factor of 2 and producing a value of the entropy that agrees
well with that of several nearby clusters.  This result is sensitive to
the thermalization of the supernova shock energy. However, as David et al
(1996) point out, the central entropy does vary from cluster to cluster,
in agreement with the scatter in the L-T plot. 

Bower (1997) has parameterized the evolution of cluster central gas
entropy as a power of the expansion factor: $s_{min}= s_{min}(z=0) +
c_{v}\epsilon ln(1+z)$ ($c_v$ is specific heat capacity of the gas at
constant volume). In this case $\epsilon=0$ corresponds to constant
entropy. Data on the evolution of the cluster XLF and L-T relationships
can then be used to constrain two parameters; $\epsilon$ and $n$, the
power spectrum index of matter density fluctuations (\cite{bow97}). Our
results imply that over the redshift range $0< z<0.4$, $\epsilon= 0\pm
0.9$. If the evolution of the ICM followed that of the dark matter
(e.g.~\cite{kai86}) then our limits on $\epsilon$ would imply $n>2$
always, thereby firmly ruling out self-similar evolution of clusters. If
there is little or no evolution in the XLF (e.g.~\cite{ebe97} ) then our
result implies that $-1.5 \ltsim n \ltsim -0.5$. Recent measurements of
the low luminosity cluster population at high redshift (\cite{jon97}) are
also consistent with zero evolution in the XLF. 

Bryan (1996) presents a scaling relation between L$_{bol}$ and T with a
$(1+z)^{3/2}$ dependence for $\Omega=1$. The dashed curve in Figure 2
represents this scaling, normalised to the luminosity band used here.
Clearly the Bryan model cannot be rejected, although it does not appear to
describe the data particularly well. The same statement can be made about
the results of Kitayama and Suto (1996) who have improved on
Press-Schechter theory by including the epoch of cluster formation, and
predict a similar (negative) temperature evolution. 

The lack of evolution in the $\sigma_V$-T relation also indicates that the
gas and the galaxies are sampling the same potential (c.f. ~\cite{car96})
and that both change in the same way with redshift, a result not generally
expected in many models. 

Presumably a similar lack of evolution would be seen in low density
universe models in which cluster evolution occurs early and structure
stops forming at $z\sim \Omega^{-1}$. However, detailed calculations for
such models are not yet available. 

\section{Conclusion}

We find no evolution in the cluster L-T relation as a function of
redshift.  These results are consistent with models in which the cluster
core acquired a high initial entropy (\cite{nav95}) as required by the
recent determination of the origin of the cluster metals (\cite{mus96})
and the lack of evolution in the cluster metallicity (\cite{mus97}).
However, it is likely that this result is also consistent with low density
cosmologies in which clusters stop evolving at relatively high redshifts,
or other variants of the standard models.  We hope that this large data
set will enourage detailed theoretical modeling. 

\acknowledgements We thank Y. Tanaka for early encouragement of this
project. This research has made use of data obtained through the High
Energy Astrophysics Science Archive Research Center Online Service,
provided by the NASA/Goddard Space Flight Center.

\clearpage

\figcaption[]{X-ray luminosity vs gas temperature for low redshift
($z<0.1$) (open squares) and high redshift ($z>0.14$) (dark triangles)
clusters. The errors in the gas temperature are symmetrized 90\%
confidence errors. The dashed and solid lines are unweighted linear
regression fits to the high and low $z$ data respectively.}

\figcaption[]{The mean cluster temperature (in the luminosity band $45.2\leq
\log_{10}L_{bol} \leq 45.7$) estimated by maximum likelihood is plotted
versus redshift. Redshift bins contain 6 objects, the highest $z$ bin at
$z\simeq 0.45$ contains 3 objects. Solid, vertical, errorbars show the 90\%
confidence limits on the mean, horizontal errorbars show size of redshift
bin. Dotted, vertical, errorbars show the intrinsic variance in T,
estimated by maximum likelihood. The dashed curve is the Bryan (1996)
model prediction normalized to this luminosity band.}

\figcaption[]{The optical velocity dispersion vs X-ray temperature. The
errors in temperature are as in Figure 1 while the errors in the velocity
dispersion are symmetrized 68\% confidence errors. We have labelled A1689
and AC118 as being obvious outliers of the distribution function. The line
drawn assumed that kT=$\sigma_V^2$ and is not a fit to the data}

\clearpage

%note all one table - split to print out in doublespace aasms4
\begin{table*}
\begin{center}
\begin{tabular}{lccl}
Name   &  Redshift & $\log_{10} L_{bol}$ & kT (keV) \\
\tableline
A1413&       0.1430 & 45.440 &         6.72($_{6.46}^{ 6.98}$)\\
A2204&        0.1530&  45.880&          8.47($_{8.05}^{8.901}$)\\
A1204&       0.1700 & 45.220 &         3.83($_{3.64}^{4.02}$)\\
A2163 \tablenotemark{a}&       0.2010&  45.600 &  12.7($_{10.7}^{14.7}$)\\
A2218&       0.1710&  45.340 &         7.04($_{7.07}^{8.01}$)\\
A586&       0.1710&  45.280 &         6.61($_{5.65}^{   7.76}$)\\
A1689&        0.1800&  45.850 &          9.02($_{8.72}^{ 9.42}$)\\
A1246&        0.1870& 45.290 &         6.28($_{ 5.79}^{   6.82}$)\\
MS0440&       0.1900&  44.970&          5.3($_{   4.45}^{6.57}$)\\
MS0839&       0.1940&  45.010&          4.19($_{  3.86}^{   4.55}$)\\
A520&       0.2010&  45.590 &          8.59($_{ 7.69}^{ 9.52}$)\\
A963&       0.2060&  45.320 &         6.76($_{  6.27}^{ 7.20}$)\\ 
A773&       0.1970&  45.500 &         9.66($_{ 8.76}^{ 10.69}$)\\
A1704&       0.2190 & 45.200  &        4.51($_{ 4.17}^{   5.07}$)\\
A1763&       0.1870 & 45.550  &        8.98($_{8.14}^{  10.00}$)\\
A2219&       0.2280 & 45.920  &       11.77($_{11.03}^{  13.03}$)\\
A2390&       0.2300 & 45.730  &        8.9($_{8.13}^{ 9.87}$)\\
MS1305+29&      0.2410 & 44.450 &         2.98($_{ 2.57}^{   3.50}$)\\
A1835&      0.2520 & 46.030 &         8.15($_{7.70}^{   8.61}$)\\
MS1455&      0.2580 & 45.480 &         5.45($_{5.17}^{   5.74}$)\\
A1758N&      0.2800 & 45.640 &        10.19($_{8.5}^{   12.48}$)\\
A483&      0.2830 & 44.900 &         6.87($_{ 5.66}^{8.46}$)\\
\end{tabular}
\end{center}

\tablenotetext{a}{\cite{mar96b}}
\end{table*}

\clearpage

\begin{table*}
\begin{center}
\begin{tabular}{lccl}
ZW3146&       0.2900 & 45.830 &         6.35($_{ 6.01}^{ 6.72}$)\\
MS1008-12&   0.3010 & 45.230 &         7.29($_{  5.77}^{ 9.74}$)\\
A1722&     0.3270 & 45.320 &         5.87($_{ 5.46}^{ 6.38}$)\\
AC118&    0.3080 & 45.810  &       12.08($_{ 11.2}^{ 13.5}$)\\
MS2137&    0.3130 & 45.440 &         4.37($_{ 4.03}^{ 4.75}$)\\
A1995&   0.3180 & 45.520 &        10.70($_{  8.9}^{ 13.2}$)\\
MS0353-36&   0.3200 & 45.270 &         8.13($_{  6.4}^{ 10.7}$)\\
MS1358&   0.3270 & 45.280 &         6.50($_{ 5.86}^{ 7.18}$)\\
A959&   0.3530 & 45.440 &         6.95($_{  5.62}^{ 8.80}$)\\
A370&   0.3730 & 45.550 &         7.13($_{ 6.30}^{ 8.18}$)\\
MS1512+36&   0.3720 & 45.050 &         3.57($_{  2.83}^{ 4.90}$)\\
A851&   0.4100 & 45.180 &         6.7($_{ 5.0}^{ 9.4}$)\\
RXJ1347-114&   0.4510 & 46.390 &        11.37($_{ 10.45}^{ 12.47}$)\\
3C295&   0.4600 & 45.380 &         7.13($_{ 5.78}^{ 9.19}$)\\
MS0451-03&   0.5390 & 45.730 &        10.17($_{ 8.91}^{  11.72}$)\\
CL0016 \tablenotemark{b}&   0.5410 & 45.600 &     8.0($_{  7.0}^{ 9.00}$)\\

\end{tabular}
\end{center}
\tablenotetext{b}{Furuzawa et al 1997, ApJ submitted}

\tablenum{1}\
\tablecomments{Numbers in brackets in final column correspond to the
90\% confidence limits on the temperatures. An inspection of other works 
(e.g. \cite{dav93}, \cite{ebe96}) indicates a scatter of $\ltsim 20$\% 
in quoted $L_{bol}$ which we attribute to differences in techniques.} 
\end{table*}

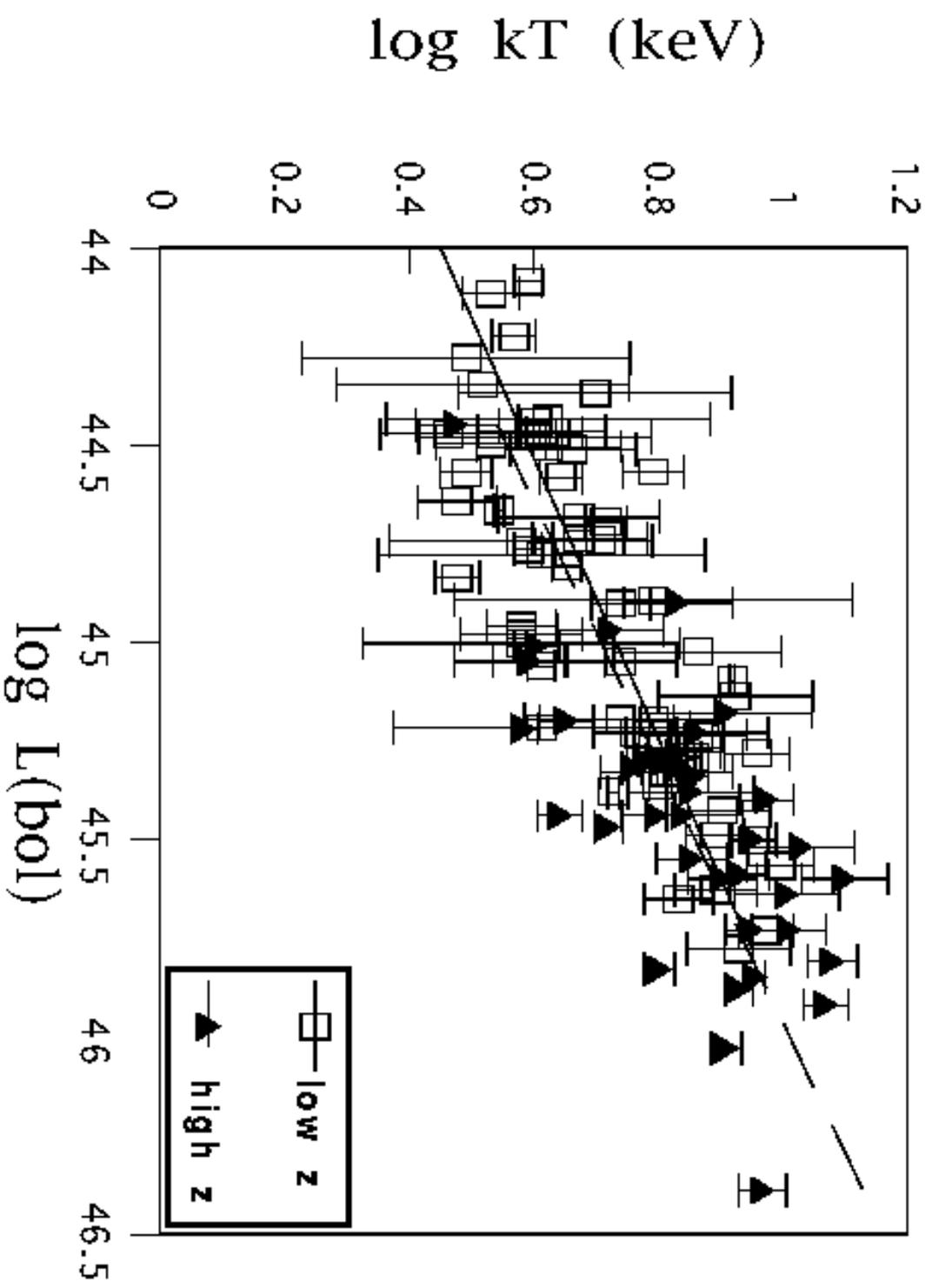
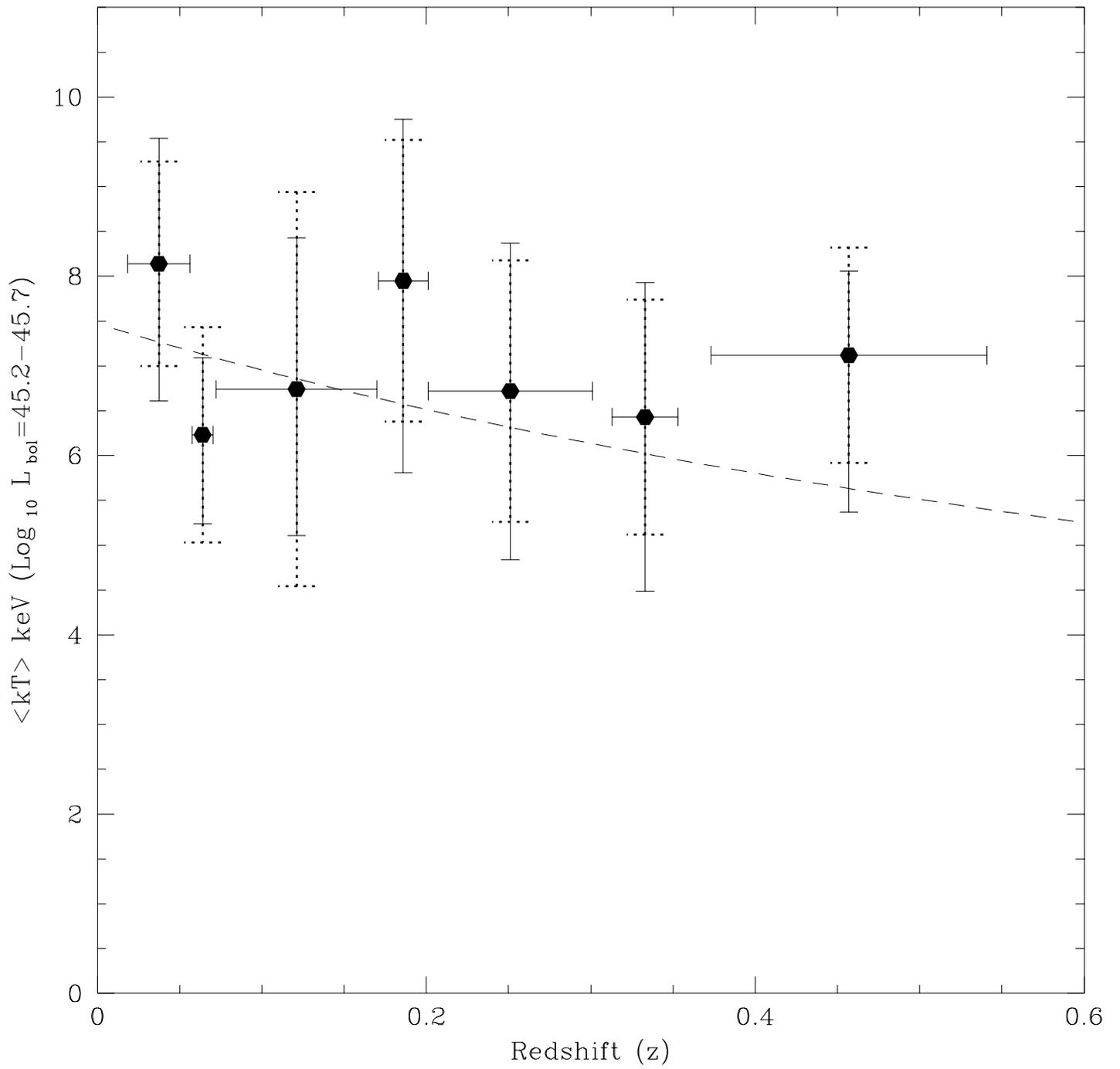
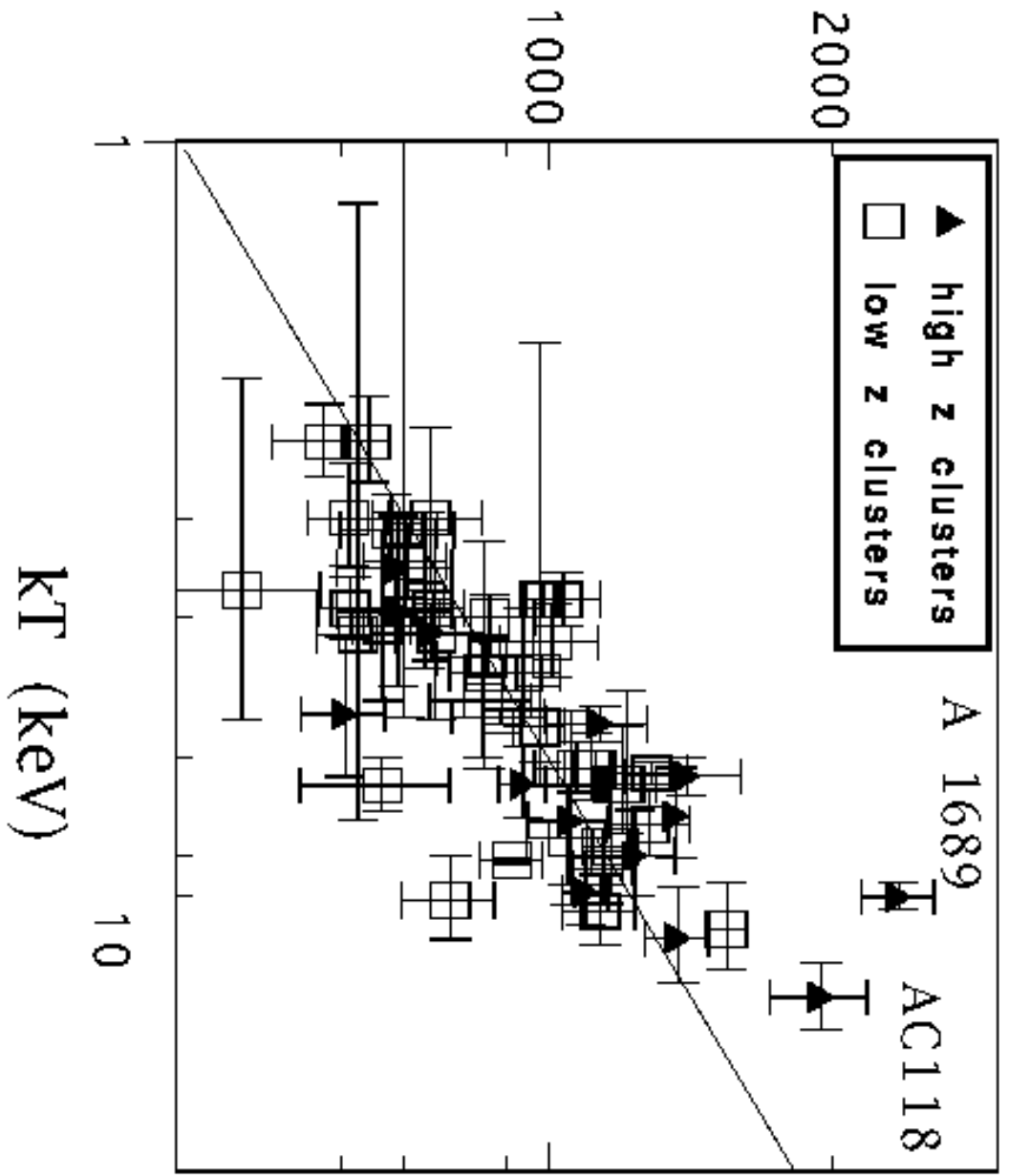
\end{document}